# Integrating Public Perspectives in Microreactor Facility Design

Diana Cambero Inda[1], Armita Marpu[1], Gina Rubio[1], Caralyn Haas[1], Prish Dhagat[1], Katie Snyder[1], Aditi Verma[1]

[1]University of Michigan, Ann Arbor, Michigan,
dianaci@umich.edu

## INTRODUCTION

Nuclear energy has been gaining momentum, driven by growing global energy demands, technical progress and supportive policies in several countries. Nuclear energy offers numerous benefits, from being a clean energy source that helps meet environmental and sustainability goals, to providing reliable base-load power that supports grid stability and ensures energy security. Rising electricity demand, led by the electrification of transport and further amplified by the rapid growth of data centers, has turned both private and governmental attention to nuclear energy as a possible solution to future energy challenges. In 2023, the global investment in nuclear energy reached approximately USD 65 billion, nearly twice the amount invested in the previous decade. This funding is being directed toward extending the lifespans of existing nuclear power plants, restarting previously shut-down reactors, advancing research and development of advanced reactor designs, and the construction of new nuclear power plants [1]. As the momentum around new nuclear energy builds, attention should also be given to the regulatory frameworks and public processes that enable their safe and legitimate deployment [2].

To bring a nuclear facility online, a formal licensing process must be followed. Each country establishes its own specific procedures but generally, licensing steps include reactor design approval, facility siting, construction authorization, and operational licensing [3]. As a part of the licensing process, each country determines the scope for public input. For example, in the Canadian licensing process, public consultation is carried out during their initial environmental assessment, followed by one or two separate public hearings during the site preparation. In the United States, public involvement typically occurs through a single hearing following the environmental and safety assessments by regulators, covering the review of the design, the site selection, and the construction plans[3]. Regardless of the approach, public and stakeholder engagement plays a significant role in achieving democratic legitimacy and efficient deployment of nuclear energy facilities [4].

Significant attention has been given to the involvement of host communities in the siting of nuclear waste repositories, particularly in countries such as Sweden, Finland, and, more recently, Canada, where distinct participatory approaches have led to successful siting outcomes. In all cases, a combination of early engagement, transparency, and voluntary participation helped foster trust in institutions and acceptance among local stakeholders [5,6]. In contrast, the United States has faced long-standing challenges in its attempts to establish a permanent repository for high-level nuclear waste. The case of the Yucca Mountain Nuclear Waste Repository illustrates how limited opportunities for public participation, non-inclusive decision-making structures, and a lack of procedural transparency contributed to widespread opposition and institutional mistrust[7,8]. These factors created a contentious environment that ultimately stalled the project and continued to complicate efforts to site a final repository in the U.S.

As interest in nuclear energy grows, the sector has the chance to build on earlier experiences and strengthen public trust through a more transparent and democratic approach that initiates public engagement as early as possible in the development of a new energy facility [2]. We believe this early-stage engagement to be vital particularly for the development of small, distributed nuclear energy facilities. To explore how public participation, particularly community engagement, might be meaningfully integrated into the design and licensing process of future nuclear projects, we facilitated a community workshop with residents of Southeast Michigan. From the workshop we gathered insights on:

a) Community ideas and suggestions for nuclear facilities and their surrounding community
b) Reflections on how community engagement could influence the facility's design
c) Values and decision-making criterias prioritized by community members
d) Community sentiments about the engagement process and the topics discussed during the workshops.

In this paper, we focus on the first set of findings: community ideas and suggestions. In doing so, we explore the following research question: what would a microreactor installation look like if it was designed with direct input from the community expected to host it?. The remaining data, including reflections on engagement impact, prioritized values, and participants' sentiment, are still being analyzed and will be presented in a forthcoming, comprehensive publication.

**Participatory design & living lab**

Community engagement in the development of public spaces is not an unprecedented practice. An example is the improvement of a challenged neighborhood in the Netherlands, facing energy poverty, low literacy, and cultural diversity. By involving the local community, students and researchers were able to design and co-create a shared living space. The authors report an enhanced sense of ownership of the community towards local challenges and strengthening of their local identity [9]. Other examples of public participation in infrastructure development include the co-design of sustainable housing with renewable energy systems with the Pinoleville Pomo Nation [10] and the creation of public spaces with integrated renewable energy technologies in a local Sweden community [11].

Participatory design is a user-centered research methodology and design practice in which the end user of a design or those directly affected by a technological innovation take on an active and iterative role as co-designers. Through their participation, users contribute their tacit knowledge, life experiences, and contextualize new designs in the real environments in which they will operate [12–14] A similar concept is that of a Living Lab. According to the European Network of Living Labs, these are "[...] open innovation ecosystems in real-life environments based on a systematic user co-creation approach that integrates research and innovation activities in communities and/or multi-stakeholder environments, placing citizens and/or end-users at the centre of the innovation process." [15] Like participatory design, living labs aim to involve end users in the co-creation of new technological solutions (ranging from product or services development and evaluation, to the design of cities and living environments), through the simulation of real-life conditions in a controlled environment, to gain insight, create higher user acceptance and stimulate innovative ideas [16,17].

**New nuclear designs in proximity to communities**

Nuclear reactor technology has existed for several decades. While safety enhancements have been made over time, most operating nuclear power plants were built before the 1980s and rely on older technology [18,19]. In recent years, significant efforts have been made to promote the development of next generation nuclear reactors. These advanced nuclear reactors promise to deliver improvements in safety, reliability and cost-effectiveness [20]. Among them are microreactors, a subcategory of small modular reactors, designed to generate less than 20 MWe and offer enhanced safety characteristics, a greater range of application, connection flexibility, and scalability. Potential applications of microreactors include resilient and reliable energy supply for remote off-grid communities, industrial locations, military bases, fast-growing megacities, and space and naval applications [21].

Due to their characteristics and intended applications, microreactors might be sited in close proximity to local communities. With the anticipated expansion of microreactors and other advanced nuclear reactors, alongside efforts to scale up large nuclear projects [1], repurposing coal power plants [22], and addressing rising interest from data centers in nuclear technologies [1], there is a growing need to adopt a more human-centered approach to nuclear project siting and design. By engaging communities directly in the design process, developers can find solutions that reflect the community needs, values, and aspirations, and lead to a smoother transition towards new energy infrastructure.

**WORKSHOP DESCRIPTION, DATA COLLECTION AND ANALYSIS**

The in-person community workshop, conducted for this study, used a participatory design approach and a living lab environment [16]. In it, 93 Southeast Michigan residents (48 community members & 45 students at the University of Michigan) were provided with both physical and virtual environments to design a hypothetical microreactor facility in their local community. The workshop methodology was adapted from Hoover et. al. and Verhey-Henke et. al.'s work [23,24]. The physical setting of the workshop encompassed in-person interactions including active learning and discussion sessions, as well as the completion of collaborative maps, and workbooks integrating community values, ideas, and concerns for the construction of a nuclear fission facility in the participants' neighborhoods. The interaction was wrapped up by the creation of a virtual prototype of the community-designed microreactor facility, using AI image generation (Fig. 3-5).

Our findings for this summary paper are drawn from the collaborative region maps. These idea maps, included four radial concentric perimeters, representing locations of a microreactor facility. Regions in the map included: "Fission Energy Facility" (innermost region), "Area to Access Facility", "Perimeter" and "Community" (outermost region). To complete this activity, participants' groups were provided with sticky notes and were instructed to independently write down ideas or suggestions they considered relevant for a nuclear facility. Afterwards, participants were asked to place the ideas in a corresponding region in the idea map (Fig. 1).

After the workshop, all collected data was scanned and transcribed into a computer-readable format. Each individual participant's idea was typed into an organized spreadsheet accounting for the location in the map where the idea was placed. A total of 1,857 ideas of community members' suggestions were collected. Two researchers and the Large Language Model ChatGPT -4o (accessed June 11, 2025), independently grouped participants' ideas into seven categories. A majority vote was used to finalize the categorization. In cases without a majority, the first author acted as a tiebreaker. If disagreement remained, the two annotators and the first author discussed the categorization until a majority vote was reached. Idea categories were taken from Hoover et. al. methodology [23]. Synthesized definitions of such categories are given here:

1. **Community Engagement:** Ways in which the community is able to interact with and the benefits that the power plant might bring (e.g. "school field trips").
2. **Aesthetics:** Related to the appearance of the facility, as well as aspects that appeal to other human senses (e.g. "minimalist design").
3. **Environmental Protection and Safety:** Maintaining the safety of the environment and the people in the community (e.g. "place in urban areas to preserve nature").
4. **Economically Beneficial:** Related to the financial feasibility and the economical benefits derived from the facility (e.g. "doesn't increase property value for the neighborhood").
5. **Communication and Ethics:** Communication methods between the facility and community members, and maintaining ethical practices and equitably meeting people's needs (e.g. "citizens must be allowed a say").
6. **Functionality/Technical:** Related to the operation, infrastructure or other technical aspect of the nuclear reactor (e.g. "solar powered microreactor", "manufacturing in available Detroit factories").
7. **Working Environment:** Accommodations, benefits, resources and working spaces for the facility's employees. (e.g. "lunch options for workers/free basic food", "day care for workers").

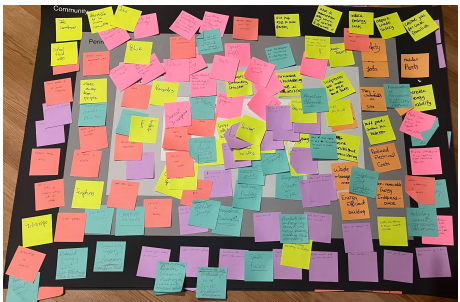

Fig. 1. Region Map scan

**RESULTS**

We found that most people's ideas (23%) for the design and siting of a microreactor facility, were related to *Community Engagement*. Some examples of these ideas are: "educational events", "day- internships to local high-schoolers", "[there] should be a science museum - free and schools can go on field trips there", "use wasted heat for community building in winter", among others. *Aesthetics* was the second most dominant category, with 22.1%. People's ideas included: "disguise it to blend with nature", "make the plant look like an art piece", "incorporate greenhouses", "fun/inviting architecture" , "transparent walls", etc. *Environmental Protection*, *Functionality/Technical,* and *Economically Beneficial* considerations were found 17.7%, 17.4%, and 12.9%, respectively. The least number of ideas involved the *Communications and Ethics* and the *Working Environment of* the hypothetical facility, with 3.7% and 3.1% representation, respectively.

Fig. 2 shows the idea distribution in each region between the nuclear facility and the surrounding community. The greatest number of ideas were directed towards the community, and the least were given for the facility itself.

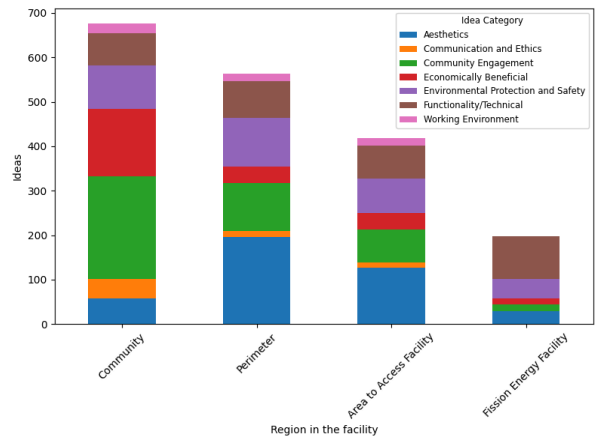

Fig. 2. Participants idea's distributed by regions in the facility.

Overall, at the *Community* level, we noticed great interest from community members in shared spaces, recreational facilities, educational and employment opportunities, and the economic (personal) savings and growth that the facility could bring (see Fig. 3 for an example of a facility design with these considerations). Environmental protection and safety ideas in this region included concerns and suggestions for: a) the siting and resource exploitation that the power plant could bring to the community, e.g.: "displaces animals and people", "made with sustainable materials", "building in an already cleared space to prevent deforestation", "built without disturbing the environment"; b) ensuring the safety of the community in case of an accident, negligent practices or produced nuclear waste, e.g.: "community safety program/ emergency alert", "does not get in our water", "limit waste as much as possible"; c) improving environmental quality by bringing a cleaner energy source, with ideas such as: "use to enrich the lives of all via not only abundant energy but a cleaner environment", "less need for coal or natural gas", "fewer people need to move to different climates", "better air quality".

For the immediate surroundings of the facility, the *Area to Access the Facility and* the *Perimeter*, people felt more strongly about the visual appearance of the buildings, with suggestions like: "no[t] obvious that its a power plant", "fits in with surrounding area", or "have a lot of walking greenery". People also included multiple ideas for the active engagement of the community in those regions. Ideas focused on the inclusion of business and commercial establishments, as well as recreational spaces such as "roller coasters", "attraction/museum/restaurant", or a "park" (see Fig. 4). In both of these regions, suggestions were made about ensuring the safety of the community and preventing any harm to the environment.

The least number of ideas were provided for the actual interior of the microreactor facility (the *Fission Energy Facility*). Some ideas were technical or functional like the incorporation of renewable energy to power the facility building; the implementation of co-regeneration processes to boost efficiency; or the use of process steam to heat-up community buildings (see Fig. 5.). Discussion in this area also addressed concerns about environmental and human safety, as well as the required safeguards to ensure the safety of the reactor.

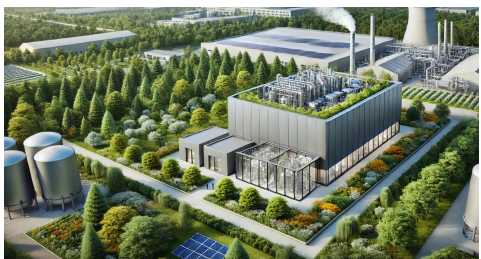

Fig. 3. Community-designed "Fission Energy Campus" microreactor facility. "A walkable facility that would encourage people to visit, increasing outreach and improving the public opinion on nuclear energy." It also includes "lots of Green Vegetation", "Research Connections and Internships" & "Partnership with the University of Michigan". [Generated with OpenAI's ChatGPT].

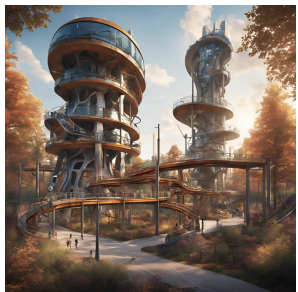

Fig. 4. Vertical community-designed microreactor facility which blends with the natural environment. Surrounding the facility, there are murals and an amusement park with a roller coaster. [Generated with Canva AI image generator].

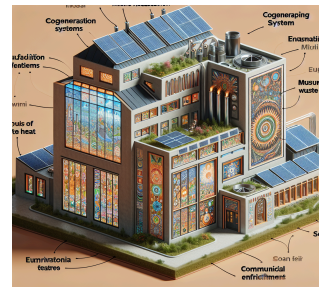

Fig. 5. Community-designed microreactor facility which, according to the designers, includes a cogeneration system to lower heating costs and improve efficiency; a museum with murals to incentivise education; solar panels for the building's sustainability; and a design that integrates with the surrounding (community) areas. [Unknown AI image generator].

Our findings indicate a desire for local infrastructure to not just provide a service (in this case, energy) but also to be a central and accessible feature of the community. Communities clearly care about socioeconomic impact, specifically about engaging with local facilities and bringing benefits and well-being to the community. Current design trajectories that call for highly automated microreactor systems requiring minimal staffing may actually be contrary to community preferences.

The limited number of ideas on the interior of the facility may suggest a deference to 'experts' on matters of technology design. Future work could focus on giving community members a deeper grounding in nuclear science and technology to see if this would elicit a larger number of ideas on the plant's technical details.

Overall, the approach presented in this work could foster meaningful conversations and benefit nuclear host communities. A recently conducted participatory design workshop, engaging young adults in the energy design decision-making process illustrates the potential application of the methodology outlined in this paper [25]. Developers should make room for community ideas to be prioritized as it may result in a trustworthy, democratic relationship with community members, which can directly impact the efficient deployment of nuclear energy facilities.